\documentclass[a4paper]{jpconf}
\usepackage{graphicx}
\usepackage{lineno}
\usepackage{booktabs}
\usepackage{subfigure}
\usepackage{changepage}
\usepackage{overpic}
\usepackage{rotating}
\usepackage{multirow} 
\usepackage{comment}
\usepackage{hyperref}

\usepackage{cite}
\usepackage{mciteplus}

\usepackage{color}
\usepackage{xspace}
\usepackage{amsmath}
\usepackage{ifthen} 

\newboolean{pdflatex}
\setboolean{pdflatex}{true} 

\newboolean{articletitles}
\setboolean{articletitles}{true} 

\newboolean{uprightparticles}
\setboolean{uprightparticles}{false} 
\newboolean{inbibliography}
\setboolean{inbibliography}{false} 

\ifthenelse{\boolean{uprightparticles}}%
{

 \def\Pmu         {\ensuremath{\upmu}\xspace}

 \def\Ppsi        {\ensuremath{\uppsi}\xspace}

 \def\PDelta      {\ensuremath{\Delta}\xspace}                 
 \def\PXi      {\ensuremath{\Xi}\xspace}                 
 \def\PLambda      {\ensuremath{\Lambda}\xspace}                 
 \def\PSigma      {\ensuremath{\Sigma}\xspace}                 
 \def\POmega      {\ensuremath{\Omega}\xspace}                 
 \def\PUpsilon      {\ensuremath{\Upsilon}\xspace}                 
 
 \def\PB      {\ensuremath{\mathrm{B}}\xspace}                 
                  
 \def\PD      {\ensuremath{\mathrm{D}}\xspace}

 \def\PJ      {\ensuremath{\mathrm{J}}\xspace}                 
 \def\PK      {\ensuremath{\mathrm{K}}\xspace}

 \def\Pi      {\ensuremath{\mathrm{i}}\xspace}

}
{

 \def\Pmu         {\ensuremath{\mu}\xspace}

 \def\Ppsi        {\ensuremath{\psi}\xspace}                 
                  
 \mathchardef\PDelta="7101
 \mathchardef\PXi="7104
 \mathchardef\PLambda="7103
 \mathchardef\PSigma="7106
 \mathchardef\POmega="710A
 \mathchardef\PUpsilon="7107
                  
 \def\PB      {\ensuremath{B}\xspace}                 
                  
 \def\PD      {\ensuremath{D}\xspace}

 \def\PJ      {\ensuremath{J}\xspace}                 
 \def\PK      {\ensuremath{K}\xspace}

 \def\Pi      {\ensuremath{i}\xspace}

}

\def\mup        {\ensuremath{\Pmu^+}\xspace}
\def\mun        {\ensuremath{\Pmu^-}\xspace} 
\def\mumu       {\ensuremath{\Pmu^+\Pmu^-}\xspace}

\def\kaon  {\ensuremath{\PK}\xspace}
  \def\Kbar  {\kern 0.2em\overline{\kern -0.2em \PK}{}\xspace}

\def\Kz    {\ensuremath{\kaon^0}\xspace}
\def\Kzb   {\ensuremath{\Kbar^0}\xspace}
\def\KzKzb {\ensuremath{\Kz \kern -0.16em \Kzb}\xspace}
\def\Kp    {\ensuremath{\kaon^+}\xspace}
\def\Km    {\ensuremath{\kaon^-}\xspace}

\def\KpKm  {\ensuremath{\Kp \kern -0.16em \Km}\xspace}
\def\KS    {\ensuremath{\kaon^0_{\rm\scriptscriptstyle S}}\xspace} 
 
\def\Kstarz  {\ensuremath{\kaon^{*0}}\xspace}

  \def\Dbar    {\kern 0.2em\overline{\kern -0.2em \PD}{}\xspace}
\def\D       {\ensuremath{\PD}\xspace}

\def\Dz      {\ensuremath{\D^0}\xspace}
\def\Dzb     {\ensuremath{\Dbar^0}\xspace}
\def\DzDzb   {\ensuremath{\Dz {\kern -0.16em \Dzb}}\xspace}
\def\Dp      {\ensuremath{\D^+}\xspace}
\def\Dm      {\ensuremath{\D^-}\xspace}

\def\DpDm    {\ensuremath{\Dp {\kern -0.16em \Dm}}\xspace}

\def\B       {\ensuremath{\PB}\xspace}
  \def\Bbar    {\kern 0.18em\overline{\kern -0.18em \PB}{}\xspace}

\def\Bu      {\ensuremath{\B^+}\xspace}

\def\Bd      {\ensuremath{\B^0}\xspace}

\def\jpsi     {\ensuremath{{\PJ\mskip -3mu/\mskip -2mu\Ppsi\mskip 2mu}}\xspace}

  \def\Y#1S{\ensuremath{\PUpsilon{(#1S)}}\xspace}

\def\L {\ensuremath{\PLambda}\xspace}

\def\Lc      {\ensuremath{\L_c}\xspace}

\newcommand{\tev}{\ensuremath{\mathrm{\,Te\kern -0.1em V}}\xspace}
\newcommand{\gev}{\ensuremath{\mathrm{\,Ge\kern -0.1em V}}\xspace}
\newcommand{\mev}{\ensuremath{\mathrm{\,Me\kern -0.1em V}}\xspace}
\newcommand{\kev}{\ensuremath{\mathrm{\,ke\kern -0.1em V}}\xspace}
\newcommand{\ev}{\ensuremath{\mathrm{\,e\kern -0.1em V}}\xspace}
\newcommand{\gevc}{\ensuremath{{\mathrm{\,Ge\kern -0.1em V\!/}c}}\xspace}
\newcommand{\mevc}{\ensuremath{{\mathrm{\,Me\kern -0.1em V\!/}c}}\xspace}
\newcommand{\gevcc}{\ensuremath{{\mathrm{\,Ge\kern -0.1em V\!/}c^2}}\xspace}
\newcommand{\gevgevcccc}{\ensuremath{{\mathrm{\,Ge\kern -0.1em V^2\!/}c^4}}\xspace}
\newcommand{\mevcc}{\ensuremath{{\mathrm{\,Me\kern -0.1em V\!/}c^2}}\xspace}

\def\ms   {\ensuremath{{\rm \,ms}}\xspace}

\def\mhz  {\ensuremath{{\rm \,MHz}}\xspace}
\def\khz  {\ensuremath{{\rm \,kHz}}\xspace}

\newcommand{\decay}[2]{\ensuremath{#1\!\to #2}\xspace}         
\def\ra                 {\ensuremath{\rightarrow}\xspace}
\def\to                 {\ensuremath{\rightarrow}\xspace}

\def\pt         {\mbox{$p_T$}\xspace}
\def\et         {\mbox{$E_T$}\xspace}

\def\gsim{{~\raise.15em\hbox{$>$}\kern-.85em
          \lower.35em\hbox{$\sim$}~}\xspace}
\def\lsim{{~\raise.15em\hbox{$<$}\kern-.85em
          \lower.35em\hbox{$\sim$}~}\xspace}

\def\BdToKstmm    {\decay{\Bd}{\Kstarz\mup\mun}\xspace}

\def\AT#1     {\ensuremath{A_T^{#1}}\xspace}           

\def\C#1      {\ensuremath{\mathcal{C}_{#1}}}                       
\def\Cp#1     {\ensuremath{\mathcal{C}_{#1}^{'}}}                    
\def\Ceff#1   {\ensuremath{\mathcal{C}_{#1}^{\mathrm{(eff)}}}}        
\def\Cpeff#1  {\ensuremath{\mathcal{C}_{#1}^{'\mathrm{(eff)}}}}       
\def\Ope#1    {\ensuremath{\mathcal{O}_{#1}}}                       
\def\Opep#1   {\ensuremath{\mathcal{O}_{#1}^{'}}}                    

\newcommand{\DKpi}{\ensuremath{\D^0\to K^-\pi^+}\xspace}
\newcommand{\DpKpipi}{\ensuremath{\Dp\to K^-\pi^+\pi^+}\xspace}
\newcommand{\Dst}{\ensuremath{D^{*+}}\xspace}
\newcommand{\DstDpi}{\ensuremath{D^{*+}\to D^0\pi^+}\xspace}

\newcommand{\BdKpi}{\ensuremath{\Bd\to K^+\pi^-}\xspace}
\newcommand{\BdDpi}{\ensuremath{\Bd\to D^0\pi^-}\xspace}

\newcommand{\BuJpsiK}{\ensuremath{B^+\to J/\psi K^+}\xspace}

\newcommand{\Jpsi}{\ensuremath{J/\psi}\xspace}

\newcommand{\etos}{\ensuremath{\epsilon^{TOS}}\xspace}
\newcommand{\etis}{\ensuremath{\epsilon^{TIS}}\xspace}

\newcommand{\etrig}{\ensuremath{\epsilon^{TRIG}}\xspace}

\begin{document}
\title{Performance of the LHCb High Level Trigger in 2012}

\author{J. Albrecht$^1$, V. V. Gligorov$^2$, G. Raven$^3$, S. Tolk$^3$}
\vskip 2mm
\address{On behalf of the LHCb HLT project}
\vskip 2mm
\address{$^1$ TU Dortmund, Germany}
\address{$^2$ CERN, Geneva, Switzerland}
\address{$^2$ NIKHEF, Amsterdam, Netherlands}

\ead{johannes.albrecht@tu-dortmund.de}

\begin{abstract}
The trigger system of the LHCb experiment is discussed in this paper
and its performance is evaluated on a dataset recorded during the 2012
run of the LHC. The
main purpose of the LHCb trigger system is to separate heavy flavour
signals from the light quark background. The trigger reduces the
roughly 11\mhz of bunch-bunch crossings with inelastic collisions to
a rate of 5\khz, which is written to storage. 
\end{abstract}

\section{Introduction}

The LHCb detector, discussed in detail in Ref.~\cite{Alves:2008zz}, is
a single arm forward spectrometer that covers a pseudo-rapidity range
of $2<\eta<5$, with the primary purpose of performing precision measurements at
the LHC. The trigger system of the LHCb experiment consists of two
levels: the first level, implemented in hardware (L0), and the High
Level Trigger (HLT), implemented in a CPU farm of about 29\,000
logical cores. The LHCb trigger system and its performance during 2011
data taking has
been described in detail in Ref.~\cite{LHCb-DP-2012-004}. This paper
describes the adjustments of the system for 2012 running and evaluates
the performance on 2012 data.

The detector includes a high-precision tracking system consisting of a
silicon-strip vertex detector surrounding the pp interaction region, a
large-area silicon-strip detector located upstream of a dipole magnet
with a bending power of about 4\,Tm, and three stations of silicon-strip
detectors and straw drift tubes placed downstream. 
Different types of charged hadrons are distinguished by information
from two ring-imaging Cherenkov detectors. Photon, electron and hadron
candidates are identified by a calorimeter system consisting of
scintillating-pad and pre-shower detectors, an electromagnetic
calorimeter and a hadronic calorimeter. Muons are identified by a
system composed of alternating layers of iron and multiwire
proportional chambers. 
The LHCb detector was originally designed for an instantaneous
luminosity of ${\cal L} = 2 \cdot 10^{32}\,{\rm cm}^{-2}{\rm s}^{-1}$. During the
running period of 2012, this design luminosity was doubled to ${\cal
  L} = 4 \cdot 10^{32}\,{\rm cm}^{-2}{\rm s}^{-1}$, which was possible thanks to
the excellent performance of both the detector hardware and the
trigger system. In these conditions, the average number of visible
interactions per bunch crossing is $\mu = 1.6$. 

The LHCb trigger system and its performance during the 2012 run will
be discussed first. Afterwards, future developments of the 
trigger system for the time after LS1 and the LHCb upgrade are
presented. 

\section{Data-driven performance evaluation}
\label{sec:tistos}

The trigger efficiency is evaluated using events that are reconstructed
using the full offline software and selected with the final analysis
selection for the respective channel. Thus, the trigger efficiency
contains only the additional inefficiency due to simplifications used
in the trigger, possible alignment inaccuracies, worse resolution than
the offline reconstruction or harder cuts imposed by rate and/or
processing time limitations. The following channels are chosen
to show the performance of the trigger: 
\begin{itemize}
\item \BuJpsiK, where the $\Jpsi \to \mumu$ channel is selected. This
  decay allows the evaluation of the muon trigger efficiency and
  serves as excellent proxy for the CP violation and rare decay
  channels containing a dimuon final state.  
\item \BdKpi as a typical two-body beauty decay.
\item \BdDpi, followed by the decay \DKpi as a typical three-body beauty decay.
\item \DKpi as a two-body charm decay.
\item \DpKpipi as a three-body charm decay.
\item \DstDpi, followed by the four body charm decay $D^{0} \ra K^{-}
  \pi^{+} \pi^{-} \pi^{+}$.
\end{itemize}
%
These channels and their selections are representative for those used
in most analyses. 
The selected charm modes cover the topologies most sensitive to CP
violating effects. 
In all off-line selected signal samples the level of
background is significantly lower than the signal. Substantial
differences in trigger efficiency, however, are observed for true
signal and background. The trigger performance on each channel is
measured by determining the signal component using fits to the
invariant mass distributions, hence avoiding any background
contamination. 

In what follows, the term “signal” refers to a combination of tracks that form the off-line
reconstructed and selected b or c-hadron candidate. To determine the trigger efficiency, trigger
objects are associated to signal tracks. The trigger records all the information needed for such an
association. All measurements of the sub-detectors have a unique identifier, and
these identifiers are written in a trigger report in the data stream for every trigger line that accepts
an event. The criteria used to associate a trigger object with a signal track are as follows:
An event is classified as TOS (Trigger on Signal) if the trigger objects that are associated with
the signal are sufficient to trigger the event. An event is classified as TIS (Trigger Independent of
Signal) if it could have been triggered by those trigger objects that are not associated to the signal.
A number of events can be classified as TIS and TOS simultaneously
($N^{TIS\&TOS}$), which allows the extraction of the trigger efficiency relative to the off-line
reconstructed events from data alone. The efficiency to trigger an event independently of the signal,
\etis, is given by \linebreak \etis = $N^{TIS\&TOS}/N^{TOS}$, where $N^{TOS}$ is the number of events classified as TOS.
The efficiency to trigger an event on the signal alone, \etos, is given by \etos = $N^{TIS\&TOS}/N^{TIS}$,
where $N^{TIS}$ is the number of events classified as TIS. 


\section{L0 Hardware Trigger} 
\label{sec:l0}


The bunch crossing frequency of the LHC is 40\mhz. This rate is
reduced by the L0 trigger to a rate of 1\mhz, at which the full
detector is read out. The L0 trigger is implemented in custom 
hardware and has a latency of 4\,$\mu$s. It triggers on high
transverse momentum (\pt) muons and on large transverse energy (\et)
deposition in the calorimeter. A relative momentum resolution of about
20\% can be reached in the L0 muon reconstruction. The thresholds
applied in L0 are given in Tab.~\ref{l0cuts}.

\begin{center}
\begin{table}[h]
\caption{\label{l0cuts}L0 thresholds in 2011 and 2012.}
\vspace{1mm}
\centering
\begin{tabular}{l@{\hspace{5mm}}*{7}{c}{c}}
\hline
&\hspace{8mm} 2011 \hspace{14mm}&\hspace{8mm} 2012 \hspace{14mm}\\
\hline
single muon&1.48\gev&1.76\gev\\
dimuon $\pt_1 \times \pt_2$ &$(1.296\gev)^2$&$(1.6\gev)^2$\\
hadron &3.5\gev&3.7\gev\\
electron&2.5\gev&3\gev\\
photon&2.5\gev&3\gev\\
\hline
\end{tabular}
\end{table}
\end{center}

\begin{figure}[b]
\begin{minipage}{18pc}
\includegraphics[width=18pc]{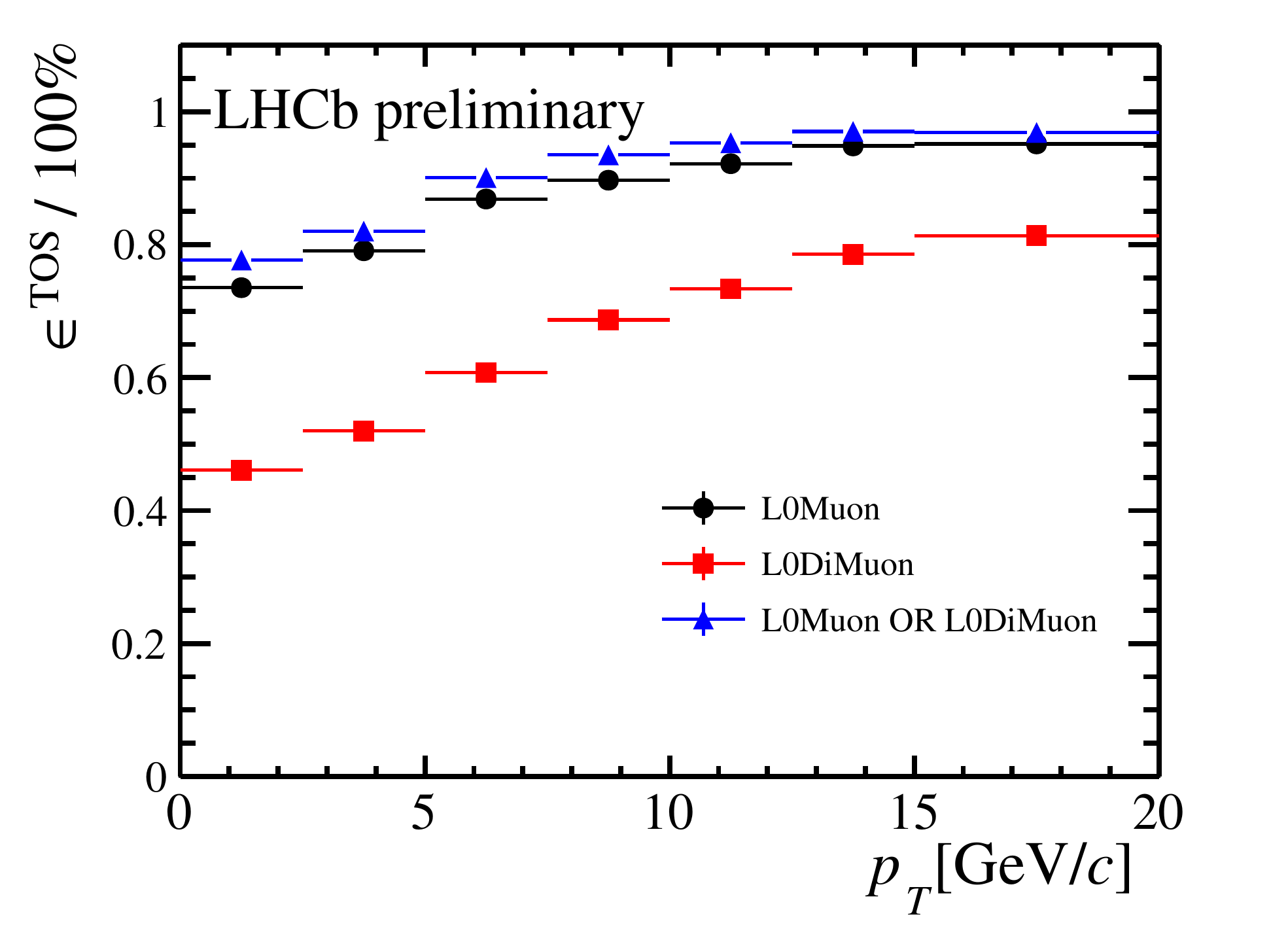}
\caption{\label{fig:l0mu} L0 muon trigger performance: TOS trigger
  efficiency for selected \BuJpsiK candidates.}
\end{minipage}\hspace{2pc}%
\begin{minipage}{18pc}
\includegraphics[width=18pc]{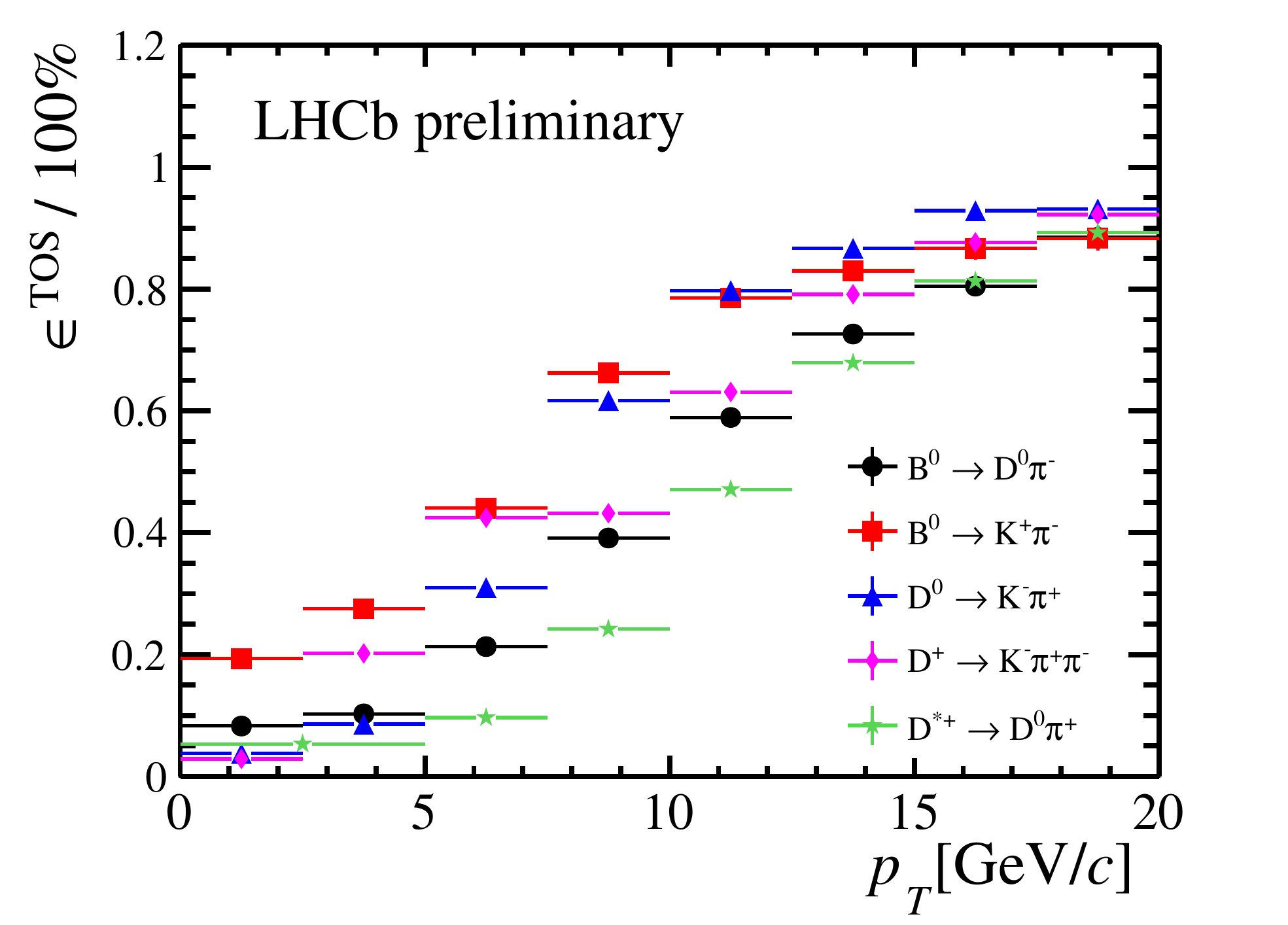}
\caption{\label{fig:l0h}
L0 hadron trigger performance: TOS trigger efficiency for different
beauty and charm decay modes.}
\end{minipage} 
\end{figure}

The trigger efficiencies are measured on offline selected events,
using the techniques described in Section~\ref{sec:tistos}. For L0
muon decisions, evaluated on \BuJpsiK events, they are shown in
Fig.~\ref{fig:l0mu}. The single muon trigger contributes the dominant
part to the efficiency. The largest inefficiency originates in the
tight muon identification requirements inside the L0 reconstruction
algorithm. The L0 dimuon trigger selects a small fraction of
additional candidates at lower transverse momenta.
The integrated efficiency for both L0 muon triggers combined is
evaluated to be 89\%.

The L0 hadron efficiency is shown in Fig.~\ref{fig:l0h} for the two
and three prong beauty decays \BdKpi and \BdDpi and the two, three and
four prong charm decays \DKpi, \DpKpipi and \DstDpi. The two prong
beauty decay is most efficiently triggered by the L0 hadron \et
criterion (\etos = 40\%) and the four prong charm decay \DstDpi
is selected with the lowest efficiency (\etos = 22\%).  The other modes
lie in between, see Fig.~\ref{fig:l0h}.
Allowing also TIS triggers, the integrated efficiencies increases
significantly, (e.g. to \etrig=53\% for \BdKpi).

The total output rate of the L0 trigger is limited to 1\mhz, the
maximum rate at which the LHCb detector can be read out. This output
rate is composed of approximately 400\khz of muon triggers (\texttt{L0Muon}
and \texttt{L0DiMuon}), 490\khz \texttt{L0Hadron} triggers and 150\khz
\texttt{L0Photon} and \texttt{L0Electron} triggers.

\section{High Level Trigger}
\label{sec:hlt}

The High Level Trigger consists of two stages, HLT1 and HLT2. The
first stage, HLT1, performs a partial event reconstruction and an
inclusive selection of signal candidates. At the reduced rate of
80\khz, HLT2 performs a full event reconstruction with only minor 
adjustments to the offline reconstruction sequence. After this
reconstruction, a set of inclusive and exclusive selections reduces
the trigger rate to 5\khz, which are saved for later offline analysis. 
The rates discussed above are average rates from the 2012 run of the
LHC, in 2011 the HLT1 output rate was approximately 40\khz and the
HLT2 output rate was 3\khz.

\subsection{First level software trigger}

The partial reconstruction in HLT1 starts by reconstructing track
segments in the vertex detector (VELO). High IP track segments and
track segments that can be matched with hits in the muon chambers are
then extrapolated into the main tracker. This extrapolation is done
using the identical forward tracking algorithm~\cite{Callot:1033584} as used in
offline processing, however, with reduced search window sizes
corresponding to  a minimum \pt requirement. 

\begin{figure}[b]
\begin{minipage}{18pc}
\includegraphics[width=18pc]{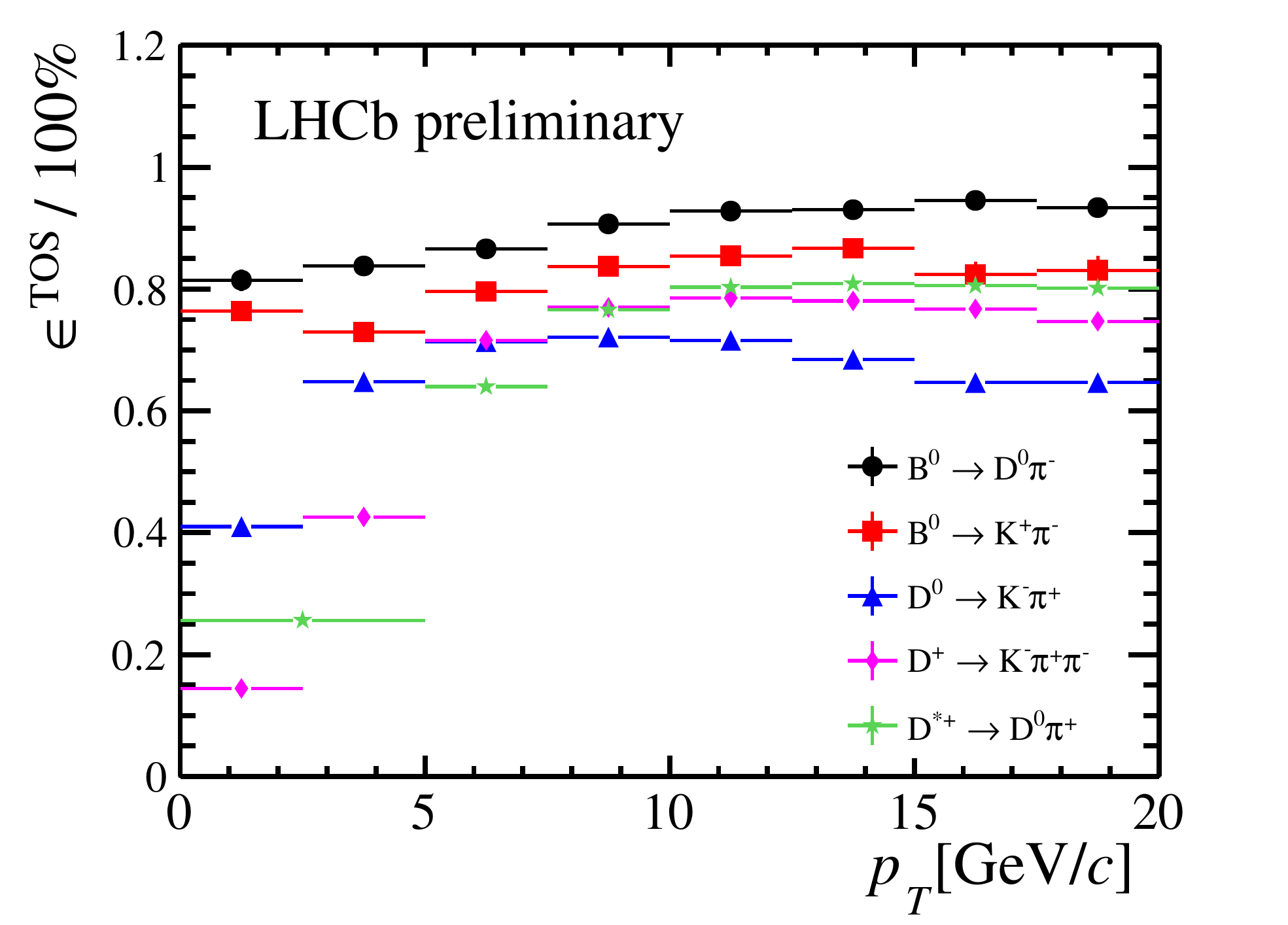}
\caption{\label{fig:hlt1h_pt} \texttt{Hlt1TrackAllL0} per\-for\-mance: TOS
  efficiency for various channels as a
  function of \B or \D \pt.}
\end{minipage}\hspace{2pc}%
\begin{minipage}{18pc}
\includegraphics[width=18pc]{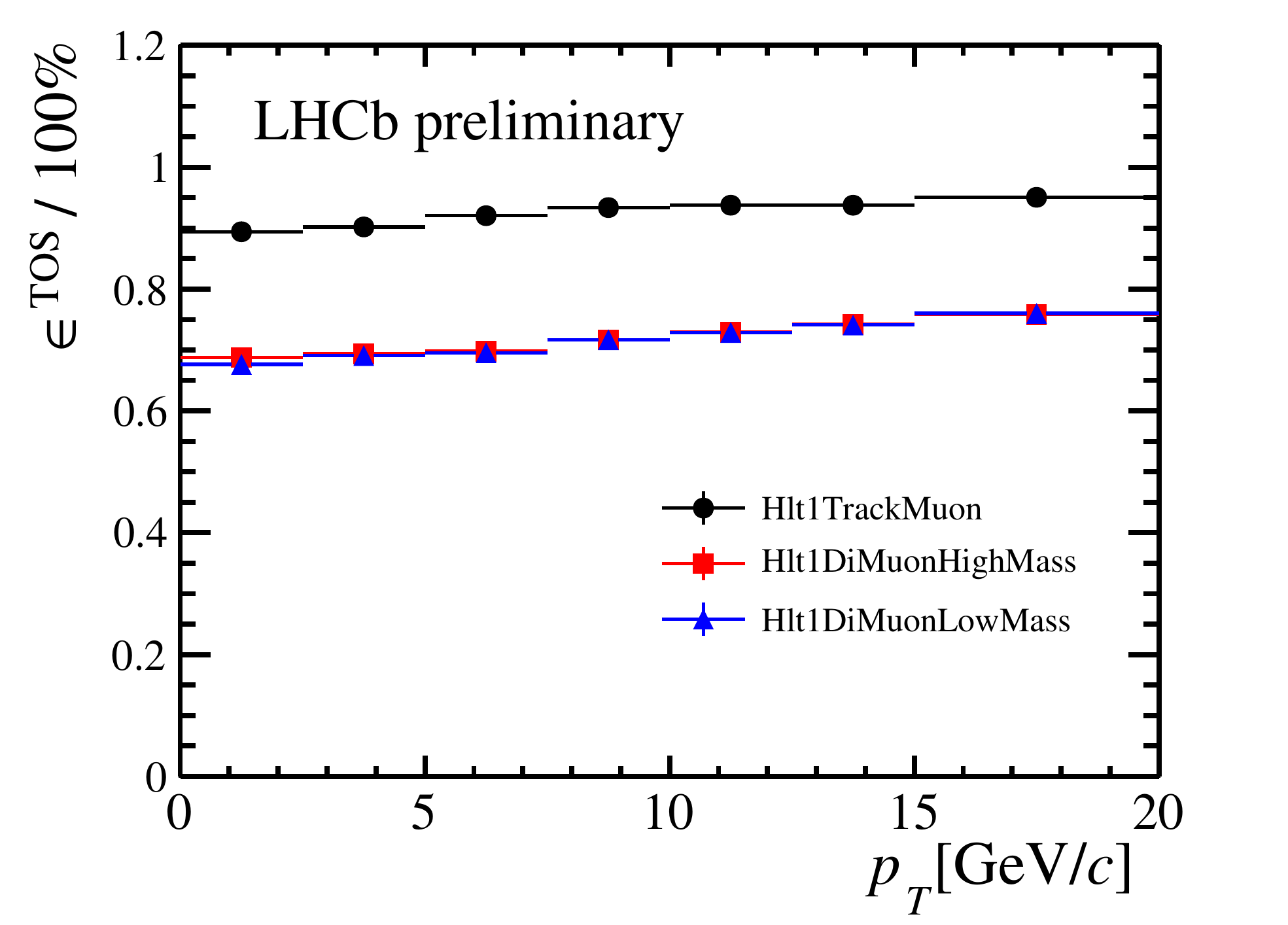}
\caption{\label{fig:hlt1mu_pt} HLT1 muon trigger performance: TOS
  efficiency for \BuJpsiK candidates as function of \Bu \pt.}
\end{minipage}%
\end{figure}

The inclusive beauty and charm trigger line \texttt{Hlt1TrackAllL0}
selects good quality track candidates based on their \pt ($\pt>1.6
\gev$) and displacement from the primary vertex. This trigger line gets the dominant part
of the HLT1 bandwidth allocated, about 58\khz. It is the dominant
trigger line for most physics channels that do not contain leptons in
the final state. The performance of HLT1 for hadronic signatures is
shown in Fig.~\ref{fig:hlt1h_pt} as a function of resonance \pt.

\vskip 2mm

A similar line exists if the track is matched with hits in the muon
chambers~\cite{Aaij:1384386}, \texttt{Hlt1TrackMuon}. This single muon trigger line
selects good quality muon candidates with a $\pt>1\gev$ that are not
coming from the primary vertex.
Single muon candidates which satisfy a \pt requirement of
$\pt>4.8\gev$ are selected by the line \texttt{Hlt1SingleMuonHighPT}
without any vertex separation requirements.

Dimuon candidates are either selected based on their mass
($m_{\mu\mu}>2.5\gev$) without any kind of displacement requirement
(\texttt{Hlt1DiMuonHighMass}), or based on the distance between
primary and secondary 
vertex, but without the mass restriction (\texttt{Hlt1DiMuonLowMass}). The dominant inefficiency for
these lines originates in the online muon identification algorithms.
The performance of HLT1 at selecting muonic signatures is shown in
Fig.~\ref{fig:hlt1mu_pt} as a function of \pt of the \Bu
candidate. The integrated efficiency is summarised in Tab.~\ref{tab:HLT1}.

\begin{center}
\begin{table}[h]
\caption{\label{tab:HLT1} Integrated efficiencies of the HLT1
  triggers, normalised to events that are offline selected (see
  Sec.\ref{sec:tistos}) and have passed the previous trigger level.}
\vspace{1mm}
\centering
\begin{tabular}{c|c|c}
\hline
channel & trigger line & \etos\\
\hline
\BuJpsiK&single muon (dimuon)&90\% (69\%)\\
\BdKpi&1Track&86\%\\
\BdDpi&1Track&89\%\\
\DKpi&1Track&67\%\\
\DpKpipi&1Track&59\%\\
\DstDpi&1Track&60\%\\
\hline
\end{tabular}
\end{table}
\end{center}

The trigger efficiency for events that are triggered by the
\texttt{L0Photon} or \texttt{L0Electron} triggers are enhanced by the
\texttt{Hlt1TrackPhoton} trigger line, which has relaxed track quality
and \pt requirements with respect to \texttt{Hlt1TrackAllL0}.

Additionally to the trigger lines discussed above, special lines are
implemented to enhance the trigger performance for events containing
candidates for high \pt electrons, di-protons, displaced vertices or
high \et jets. A set of technical lines including selections for
luminosity and beam gas measurements complete the list of HLT1
triggers.

\subsection{Second level software trigger}

The second software trigger level, HLT2, performs a full event
reconstruction for all tracks with a minimum \pt of 300\mev. It
reduces the event rate to 5\khz, which is written 
to permanent storage. 
Several exclusive and inclusive selections are performed in this
trigger level, the most important ones are discussed in this section. 

\begin{figure}[h]
\begin{minipage}{18pc}

\includegraphics[width=16pc, height=13pc]{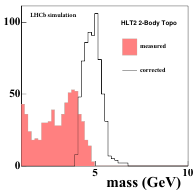}
\caption{\label{fig:mcorr} Simulated \BdToKstmm events: reconstructed
  2-body mass in red and corrected mass (see text for definition) in black.}
\end{minipage}\hspace{2pc}%
\begin{minipage}{18pc}
\includegraphics[width=18pc]{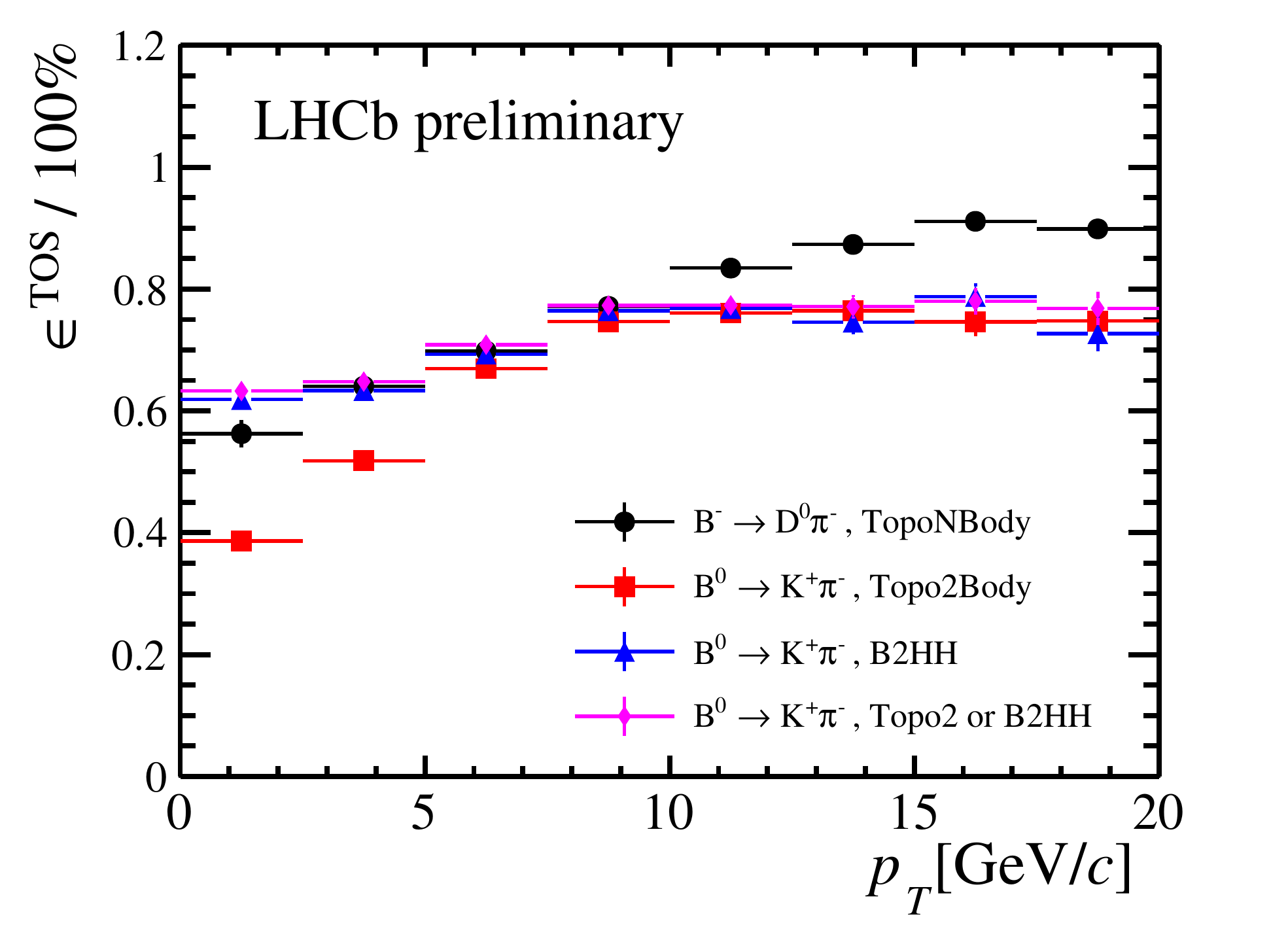}
\caption{\label{fig:topo} HLT2 inclusive beauty trigger performance as
a function of \B \pt. The efficiency for the exclusive \BdKpi trigger
line is also given.}
\end{minipage}%
\end{figure}

\vskip 2mm
\noindent\textbf{Generic beauty trigger}\\
A multivariate selection is used to trigger \B decays into charged
hadrons in an inclusive selection based on two- three- and four-prong
vertices~\cite{Gligorov:1384380}. These trigger lines, named \texttt{Hlt2Topo(N)Body}, are
based on a BDT classifier that uses discretized input
variables~\cite{bonsaiTree} which ensures a fast and robust implementation. A
crucial input to the BDT is the corrected mass, defined as
\begin{equation}
m_{corr}=\sqrt{m^2+|p_{T}^{miss}|^2}+p_{T}^{miss}\, ,
\end{equation} 
where $p_{T}^{miss}$ denotes the missing momentum transverse to the
direction of flight. This allows the topological trigger to select
heavy flavour decays even in case not all final state particles are
reconstructed. Fig.~\ref{fig:mcorr} shows the reconstructed 2-body
mass for \BdToKstmm events, and superimposed, the corrected
mass. Fig.~\ref{fig:topo} shows the efficiency for the topological
trigger lines for \BdKpi and \BdDpi events as well as the additional
efficiency that can be gained by an exclusive selection for \BdKpi in
the low \pt regime. The output rate of the topological trigger is
2\khz.  
\begin{figure}[h]
\begin{minipage}{18pc}
\includegraphics[width=18pc]{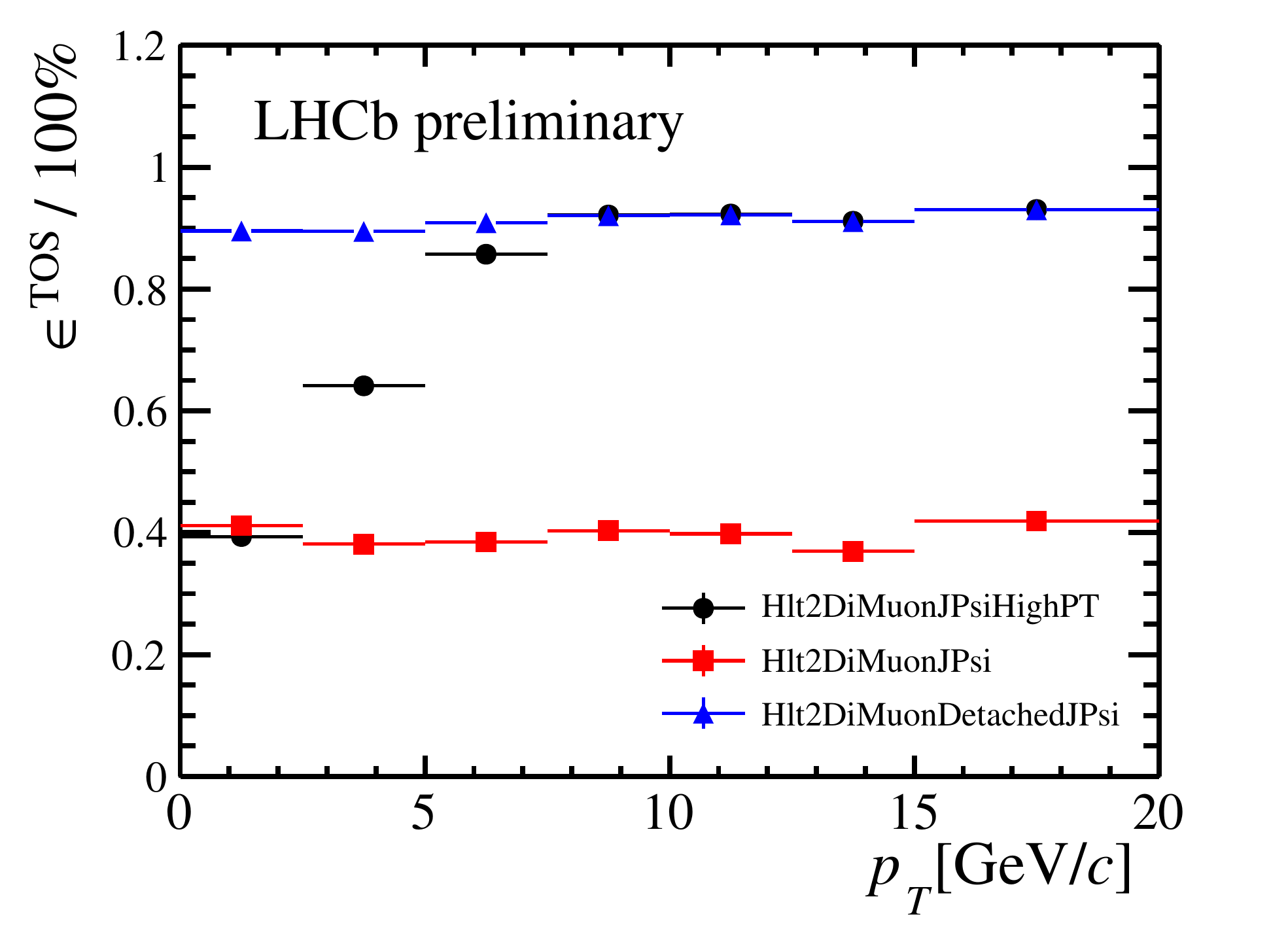}
\caption{\label{fig:hlt2mu} HLT2 muon trigger performance for the
  \jpsi trigger lines.}
\end{minipage}\hspace{2pc}%
\begin{minipage}{18pc}
\includegraphics[width=18pc]{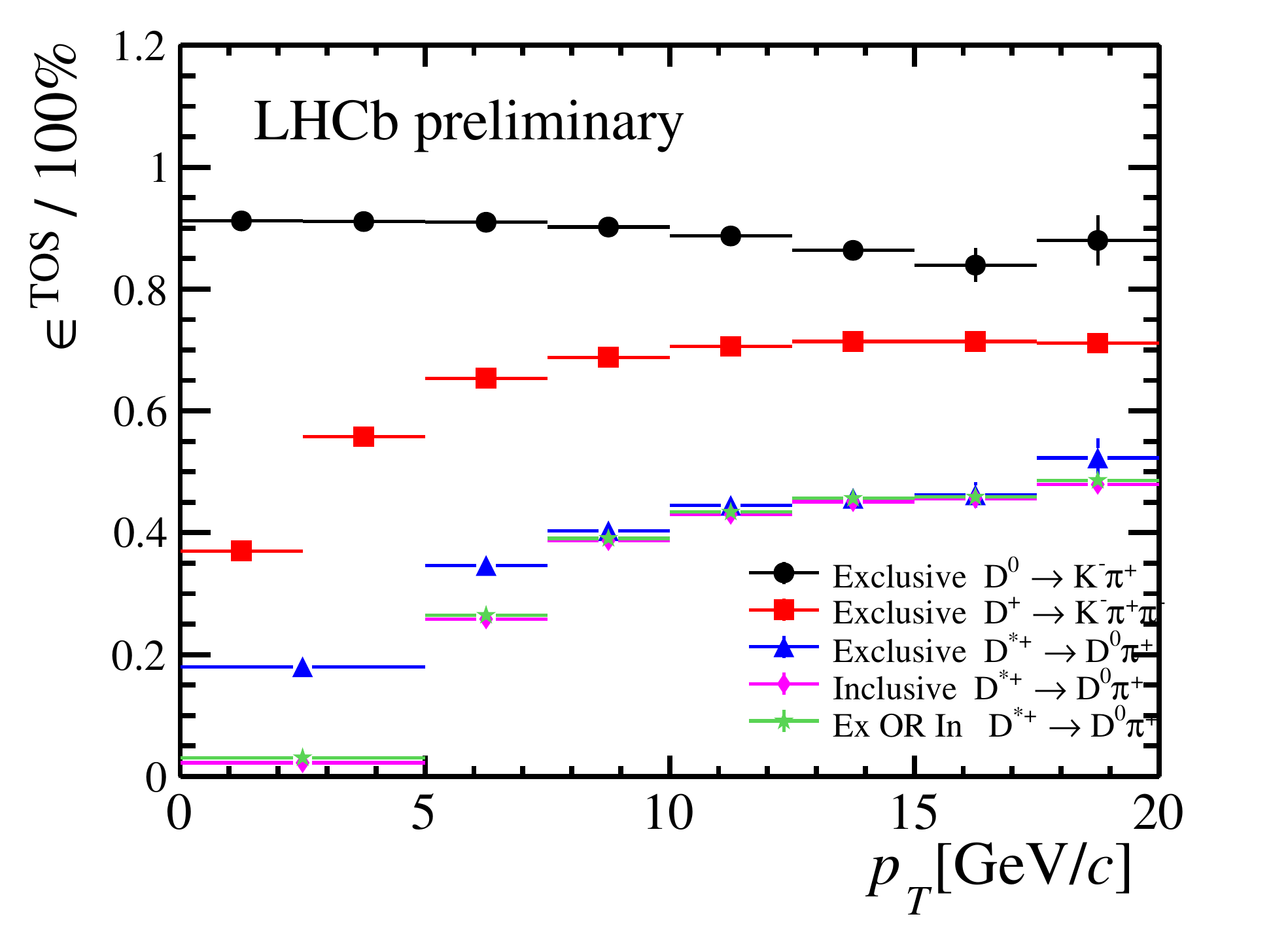}
\caption{\label{fig:hlt2c} HLT2 charm trigger performance for
  inclusive and exclusive selections.}
\end{minipage}%
\end{figure}

\vskip 2mm
\noindent\textbf{Muon triggers}\\
Several trigger lines select events with one or two identified
muons. The muon identification procedure in HLT2 is identical to the
one used in offline analysis~\cite{LHCbnote:muonid}. Single muon
candidate events are selected if the muon passes a tight \pt requirement
($\pt>10\gev$) or if candidates not consistent with the primary vertex
as their origin and pass moderate \pt ($\pt>1.3\gev$) and tight
track quality requirements. Additionally, the latter line is pre-scaled
by a factor of two.

Dimuon candidate events are selected without a mass requirement if the
dimuon vertex is separated from the primary vertex. If the mass of the
muon pair is consistent with a \jpsi resonance mass, three trigger
selection are applied
\begin{itemize}
\item \texttt{Hlt2DiMuonJPsi}: All candidates in a mass range of
  100\mev ($\approx 8\sigma$) around the \jpsi mass are accepted. In
  a part of the dataset, this prompt \jpsi trigger was pre-scaled.
\item \texttt{Hlt2DiMuonJPsiHighPT}: If the \jpsi resonance \pt is
  above 2\gev, the event is selected.
\item \texttt{Hlt2DiMuonDetachedJPsi}: If the \jpsi candidate vertex
  is separated from the primary vertex, the event is triggered. 
\end{itemize}
This set of lines is optimised to fully exploit the large physics
potential in both prompt \jpsi and $\B \to \jpsi X$
decays. Fig.~\ref{fig:hlt2mu} shows the performance of the \jpsi
triggers. The effective pre-scale of about a factor two on the prompt
\jpsi line \texttt{Hlt2DiMuonJPsi} is visible, as well as the \pt turn
on of the \texttt{Hlt2DiMuonJPsiHighPT} line.  
The total output rate of all single and dimuon trigger lines is about 1\khz.

\vskip 2mm
\noindent\textbf{Charm triggers}\\
In the 2012 running, about 600\khz of $c\bar{c}$-events are produced
in the acceptance of the LHCb spectrometer. This high rate implies
tight cuts on the invariant mass in exclusive trigger selections. Only
the decay chain \DstDpi can be selected inclusively, i.e. only
reconstructing two charged tracks from the \Dz decay matched to a slow
pion from the \Dst decay. 
The mass difference between the \Dst and \Dz candidates remains a good
discriminating variable because of the small q-value of the decay,
enabling the rate to be sufficiently reduced.
The \Dz is partially reconstructed in all
different combinations of $\pi^\pm$, $K^\pm$, $p$, $\mu^\pm$, \KS or $\Lambda^0$ 
allowing both rare decay and CP violation measurements.
The dominant exclusive selections for prompt
charm are the hadronic two body selection \texttt{Hlt2CharmHadD02HH}
and three body selection \texttt{Hlt2CharmHadD2HHH}. The efficiency of
these trigger lines is summarised in Fig.~\ref{fig:hlt2c}. 
Further selections for hadronic, leptonic and semi-leptonic \D and \Lc
decays are implemented. 
The total output rate of all charm selections is about 2\khz. 


\section{Deferred trigger}

The LHC machine delivers stable beams only for about 30\% of the time. In
order to use the idle time of the Event Filter Farm, the local hard
discs of the Event Filter Farm nodes are used. In this approach, called deferred
triggering, about 20\% of the L0 accepted events are temporarily
saved. They are analysed at a later time during the inter-fill
gaps. The scheme of deferred triggering is discussed in more detail in
Ref.~\cite{mFrank}. This more optimal usage of CPU resources allowed
to improve the track reconstruction in HLT2 in two ways:
\begin{itemize}
\item
it allowed to decrease the \pt requirement inside the forward tracking
algorithm from 500\mev to 300\mev, and 
\item
allowed the implementation of a specialised track reconstruction for
long lived particles (e.g. \KS), which reconstructs tracks also if no
hits are found in the Velo detector. 
\end{itemize}
These modifications of the trigger configuration enhanced the
performance to select charm decays significantly.

\section{Future developments beyond LS1}

The restart of the LHC accelerator after long shutdown 1 is
planned for 2015. The HLT1 and HLT2 will then be decoupled from each
other and  the deferral of events will occur after they are accepted
by HLT1. This deferral at a reduced rate allows to buffer the data
significantly longer and thus allows to perform an online calibration
of the detector. 
The most important part here is the run-by-run calibration of the
refractive index and the mirror alignment of the RICH detectors. 
This will allow PID selections in HLT2 which are very close in
performance to the currently used offline selections. 
This enables to specifically select Cabibbo suppressed charm decays
and suppress the dominant Cabibbo favoured decays with pre-scales.
Details of the planned splitting between HLT1 and HLT2 are
given in Ref.~\cite{sebastian}.
In addition, LHCb will be able to profit
from a larger trigger farm. Increased computing resources available in
2015 will allow to record about 12.5\khz of events to disk. 
\vskip 2mm

\begin{figure}[h]
\centering
\begin{minipage}{18pc}
\includegraphics[width=18pc]{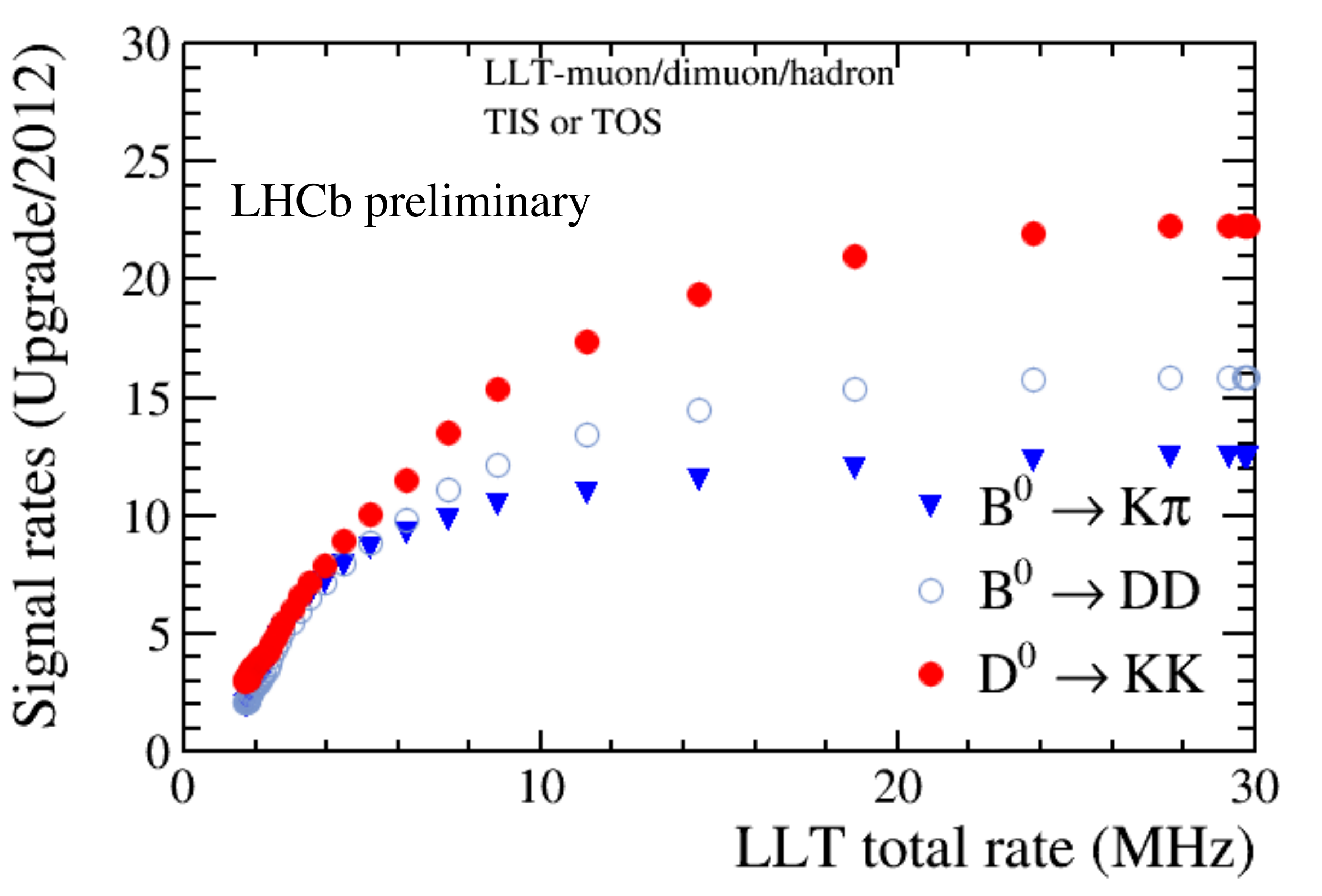}
\caption{\label{fig:upgrade} Expected increase in signal yield per
  unit luminosity dependent on the LLT output rate for three benchmark
channels.} 
\end{minipage}\hspace{2pc}%
\end{figure}
The upgraded LHCb detector following long shutdown 2 will
allow a detector readout at the full bunch crossing rate of 40\mhz. 
A Low Level Trigger (LLT) with variable output rate between 1 and 40\mhz
will initially reduce the HLT input rate until the full DAQ network
and CPU farm are installed. 
Fig.~\ref{fig:upgrade} shows the factor in signal yield for hadronic
channels that can be gained if the 1\mhz L0 trigger bottleneck is
removed by a  a flexible LLT. From LLT rates of 10\mhz onwards,
factors between 10 and 20 can be gained, depending on the particular
decay channel.

\section{Summary}
The LHCb trigger system has been performing extraordinarily well in the
first running phase of the LHC. 
It is designed to select charm and beauty hadrons in a large range of decay
modes and permits the measurement of its efficiency directly on data. 
The flexible design of  the HLT, fully
deployed in software, allows to quickly adjust to changes in running
conditions and physics goals. Inclusive selections in the full trigger
chain allow an efficient trigger for basically any beauty decay to
charged tracks. 
Several innovative concepts have enabled this performance: the
deferred triggering allows to optimise the trigger usage for mean
instead of peak usage of the available computing
resources. Multivariate selections allow the inclusive selection of
beauty decays into charged tracks with high efficiency. 

For the future, several improvements of the trigger system are
planned: the two software trigger levels will be decoupled which
allows calibrations between them and thus a performance much closer to
the one achieved offline.  
LHCb is preparing to upgrade the detector in 2018. It will feature a
fully software based trigger system that will allow to explore the
physics goals of the experiment at significantly increased
luminosities.





\section*{References}
\addcontentsline{toc}{section}{References}
\setboolean{inbibliography}{true}
\bibliographystyle{LHCb}
\bibliography{main,LHCb-PAPER,LHCb-CONF,LHCb-DP,article}


\end{document}